\documentstyle[12pt,epsfig]{article}
 \title{
Channeling of high-energy particles in a multi-wall nanotube}

\author{ S. Bellucci$^a$\footnote{Corresponding author, bellucci@lnf.infn.it}, V.M. Biryukov$^b$, A. Cordelli$^a$ \\
\small \it
$^a$  INFN, Laboratori Nazionali di Frascati, Via E. Fermi 40, 00044 Frascati, Italy\\
\small \it
$^b$ Institute for High Energy Physics, 142281 Protvino, Russia
}
\date{}
\begin{document}
\baselineskip=23pt
\maketitle

\abstract{
\normalsize
\baselineskip=23pt
Channeling of high-energy particles in straight and bent multi-wall nanotubes (MWNT) has been studied in
computer simulations and compared to the channeling properties of single-wall nanotubes (SWNT) and bent crystal
lattices. It is demonstrated that MWNT can efficiently channel positively-charged high-energy particles trapped between
the walls of MWNT. Bending dechanneling in MWNT has been computed as a function of the particle momentum to
nanotube curvature radius ratio, $pv/R$.
It is found that a bent MWNT can steer a particle beam with bending capabilities similar to
those of bent silicon crystal lattice and to those of best (i.e. the narrowest) SWNT.
In view of channeling applications at particle accelerators,
MWNT appear favored as compared to SWNT, because MWNT can be produced quite straight (and in aligned array),
while SWNT is typically very curved, thus posing a severe problem for channeling applications.
Therefore, we suggest that MWNT provide a better candidate for channeling than SWNT.
}
\pagebreak

\section{
Introduction}

Bent channeling crystals are well established as a technical tool for steering  of particle beams at accelerators [1].
The related physics has been experimentally tested in the energy range
spanning over six decades, from 3 MeV \cite{2} to 900 GeV \cite{4,5}. Today, bent crystals are broadly
used for extraction of 70-GeV protons at IHEP (Protvino) with efficiency of 85\% routinely obtained at intensity well
over 10$^{12}$ particle per second \cite{7}.
Channeling technique is used at IHEP for beam delivery since 1987 on everyday basis \cite{8,13}.
About ten channeling crystals are installed on six locations in the main ring of the 70 GeV proton
accelerator of IHEP.
A bent crystal (5 mm Si) was installed into the Yellow ring of the
Relativistic Heavy Ion Collider where it channeled Au ions and polarised protons of 100-250 GeV/u,
within the framework of the collimation experiment \cite{16}.
There has been interest to apply channeling technique at accelerators from TeV colliders for collimation and
extraction \cite{19,20,21,22} down to sub-GeV microbeam facilities \cite{23}, or for instance to use it for a channeling
undulator \cite{24,251,25}.

Nanostructured material offers an interesting alternative to crystal lattices as a guide for channeled
particles \cite{26,261}.
First, it is a material with very unusual properties \cite{34}. The channels in the nano-ordered material can be
much wider than those in crystals, and the channeled particles are trapped in two dimensions (in a nanotube)
rather than in one dimension like with crystal planar channels.
Second, the nano-material can be designed to fit applications the best, in principle, with a wide choice of
geometrical characteristics and atomic content \cite{34,341}.

Single-wall nanotube (SWNT) channeling has been studied in computer simulations by several authors \cite{26,36,36,37}.
In our previous simulation study \cite{38} of channeling in straight and bent SWNT we answered the question of
what kind of nanotube geometry (diameter) would fit best the application of SWNT for particle beam steering at
accelerators.
Another essential question answered \cite{38} was: how SWNT compares to bent crystal lattice in
particle channeling.
Narrow (order of 1 nm in diameter or less) nanotubes were found to have an efficiency of
beam bending similar to that of silicon crystal lattice, while wider nanotubes appear useless for particle
steering because of poor efficiency \cite{38}.

Multi-wall nanotubes (MWNT) have not been a subject of channeling research so far. While SWNT with its wide channel
looks at first glance a naturally attractive object for channeling research, MWNT is a rather dense pack of walls
with the spacing of about 0.34 nm (in the case of carbon) which is much narrower than the width of SWNT. However, it is
still about two times larger than the spacing in the crystal lattice channels.
MWNT are a very common nanostructure and actually were discovered first, ahead of SWNT.

An essential feature of MWNT is that it is very straight on production, unlike SWNT which is very curved.
Also, MWNT are much more easily produced as an aligned array of straight parallel tubes than SWNT.
From the practical viewpoint of channeling applications all this is of paramount importance.
However, the channeling properties of MWNT have to be demonstrated first and clarified.
Further on, the fact found in the study \cite{38} that the spacing in the channel should not be too large,
in order to make it efficient has also stimulated us to try MWNT for channeling.
Finally, without Monte Carlo simulation of a particle beam interaction with a nanostructure, it is very
difficult to decide what kind of nanostructure would be best suited for channeling.

\section{ Channeling in MWNT}

For Monte Carlo simulations of particle interaction with a MWNT we applied the same model as in our previous
work \cite{38} on SWNT channeling. The model was upgraded to take into account multiple walls.
A MWNT has an internal diameter (the one of the narrowest tube in the pack) and
an external diameter. Their typical values are a few nanometers and a few tens of nanometer
respectively, for carbon nanotubes \cite{34,341}.

Channeling inside the internal diameter of MWNT is not of interest, because of poor efficiency shown
due to a large width of this channel \cite{38} and because of its small cross-section with respect to the total
cross-section of the MWNT. Therefore, we were interested in channeling between the walls in the bulk of MWNT.
Our simulations were done for carbon MWNT.

As previously, we average the potential $U(\rho,\phi,z)$
of a straight nanotube over the longitudinal coordinate $z$ and azimuth angle $\phi$ to obtain a
potential $U(\rho)$ with cylinder symmetry.
As in crystal channeling, the averaging over $z$ is well justified as a collision of a particle with a nanotube wall
under a glancing angle does involve many thousand atoms along the particle trajectory. For the same reason, the
averaging over $\phi$ is equally justified if the nanotube has an arbitrary helicity \cite{36}
as defined by nanotube indices ($m, n$).
In the special cases of zigzag ($m=0$) or armchair ($m=n$) nanotubes, the wall consists of atomic rows parallel to the
nanotube axis; the nanotube potential is then defined by the sum of potentials of the rows, and this case
deserves a separate consideration.
Further on in the Letter we
apply only the averaged potential $U(\rho)$ for a straight nanotube.
This approach reveals the general features of nanotube channeling.

In a carbon SWNT, the channeled particles are confined in a potential well
$U(\rho)$ with the depth $U_0$ of about 60 eV.
The field experienced by a high-energy particle moving in an aligned MWNT is the sum of the fields
of single walls of different diameter with the same axis.
In addition to atomic potentials, in a nanotube bent with radius $R$,  an
effective potential taking into account a centrifugal term $pvx/R$ is introduced
similarly to bent crystals [1]:
$U_{eff}(\rho,\phi)=U(\rho,\phi)+pvx/R$, where $x=r \cos(\phi)$ is the coordinate in the
direction of bending, $pv$ is the particle momentum times velocity.

We use so-called standard potential introduced by Lindhard \cite{39}. When averaged over ($\phi, z$),
the potential of SWNT
is described by \cite{26}:
\begin{equation}
U_{SWNT}(\rho )=\frac{4NZ_1Z_2e^2}{3a}\ln \left(\frac{r^2+\rho^2+3a_S^2+\sqrt{(r^2+\rho^2+3a_S^2)^2-4r^2\rho^2}}
{\mid r^2-\rho^2\mid +r^2+\rho^2}\right)
 \end{equation}
Here $Z_1e, Z_2e$ are the charges of the incident particle and the nanotube nuclei respectively, $N$ is the number of
elementary periods along the tube perimeter, $a$=0.142 nm is the carbon bond length;
the SWNT radius is $ r=Na\sqrt{3}/2\pi$ .

The screening distance $a_S$ is related to the Bohr radius $a_B$  by
\begin{equation}
a_S=\frac{a_B}{2} \left(\frac{3\pi}{4(Z_1^{1/2}+Z_2^{1/2})}\right)^{2/3}
\end{equation}
for interaction between neutral atoms. In the D. Gemmel's review \cite{gemmel} this formula is applied
also to partially ionized projectiles. Further discussion of screening distance can be found in ref. \cite{gemmel}.
For a point-like charge or fully ionized projectile, a simpler formula is used:
\begin{equation}
a_S=\frac{a_B}{2} \left(\frac{3\pi}{4}\right)^{2/3}Z_2^{-1/3}
\end{equation}

The potential $U(\rho)$ of MWNT is the sum of contributions of SWNT making up the MWNT. Actually, only the two adjacent
walls contribute sizably for the particle located between them.

In a tube bent along the $x$ direction, the motion of a particle is described by the equations
\begin{equation}
pv\frac{d^2x}{dz^2}+\frac{dU(\rho)}{dx}+\frac{pv}{R(z)}=0
 \end{equation}
\begin{equation}
pv\frac{d^2y}{dz^2}+\frac{dU(\rho)}{dy}=0
 \end{equation}
where $\rho^2=x^2+y^2$. This takes into account only the nanotube potential and the centrifugal potential.
Any particle within close distance, order of $a_S$, from the wall (where density of the nuclei is significant) is also
strongly affected by the nuclear scattering.

Two mechanisms of particle transfer from channeled to random states are well known for crystals: scattering on
electrons and nuclei and curvature of the channel \cite{1}. A typical nanotube is less than 0.1 mm in length at present.
For such a short channel, the scattering on electrons within the bulk of the tube is insignificant for high-energy
particles. However, a curvature of the tube could quickly (in less than one oscillation) bring much of
the channeled particles out of the potential well or
into close collisions with the nuclei of the nanotube walls.

Figure 1 shows (a) an example of the trajectory of a particle channeled in a straight MWNT, trapped between
the pair of adjacent walls. The radial motion of the channeled particle is finite while in azimuthal direction
it is free. Figure 1 (b) shows an example of the trajectory when MWNT is weakly bent; obviously, the particle is
localised in some range of  $\phi$ to conform to a centrifugal force when moving along a bent nanostructure.
An example of the stronger bending is illustrated by Figure 1 (c).

As a result of bending, the MWNT phase-space of transverse coordinates and transverse angles available for
channeled particles is reduced. The particles channeled through a bent MWNT are deflected at the angle of
MWNT bending.
Figure 2 shows an example of the angular distribution of particles downstream of the 50-$\mu$m long MWNT
bent 5 mrad, shown in the direction of bending. Similarly to the pictures of bent crystal channeling, there
is clear separation of channeled and nonchanneled peaks, with some particles lost (dechanneled along the tube)
between the peaks.

This example shows that channeled particles can survive in a bent MWNT, so the effect could be used
for steering of high-energy particles provided its efficiency in MWNT is good enough compared to SWNT
and crystal lattices.

\section{ Comparison of MWNT to SWNT and to crystal lattice}

While the capability of MWNT to channel particles could be hoped for from the standpoint of channeling theory,
just from the existence of channels, the channeling efficiency of MWNT relative to other channeling structures
such as SWNT and crystal lattices is not obvious at all, unless this issue is studied in Monte Carlo simulations.

We looked in simulations how channeled particles survive in multi-wall nanotubes of different curvatures.
The Monte Carlo studies of nanotube channeling efficiency were done over a broad range of bendings for the nanotube,
and with a range of MWNT size (inner and outer diameters) about its typical value in a synthesized MWNT.
A parallel beam of 1 GeV protons was entering a carbon nanotube, where protons were tracked over 50$\mu$m.
Multiple scattering was not included, so we did evaluate only the effects of bending dechanneling.
Figure 3 shows the number of protons channeled through 50$\mu$m of MWNT as a function of the
centrifugal force $pv/R$; for comparison, we also show the same function for SWNT and Si (110) crystal lattice.

Figure 3 shows similar slopes of the curves for MWNT, SWNT and Si(110), this means that
the number of channeled particles in a MWNT declines at a rate similar to that in the other efficient
channeling structures. This holds true both for the moderate bendings 0.1 to 1 GeV/cm (equivalent to 30-300 T
magnetic field, Figure 3 (a)) where silicon channeling crystals are used
for beam steering at the high-energy accelerators nowadays,
and for the strong bendings of 1 to 4 GeV/cm (equivalent to 300-1200 T, Figure 4 (b)).
Such a study was done for MWNT of different external and internal diameter, and the same results were obtained;
the parameter that matters was the interwall spacing fixed at 0.34 nm.
Notice that MWNT competes with the best of SWNT ($\leq$1 nm diameter) while wider SWNT ($\ge$1 nm) are less efficient
in steering (bending) of the channeled particles.

\section{ Further research and potential applications}

The feasibility of experiments to test channeling and coherent scattering with the existing samples of very
short (tens of micron) nanotubes at high energy accelerators has been demonstrated
in Monte Carlo simulations \cite{40,401}.
These studies could be extended also to the comparison of MWNT versus SWNT.
Such experiments are in preparation at IHEP and LNF. Practical considerations,
such as the demand for a good alignment within the array of nanotubes, could play a leading role in the choice
of nanostructured material for channeling studies. Since we have shown that MWNT are at least as good channeling
structures as SWNT, further considerations driven by technology may well switch the general interest from SWNT
to MWNT as channeling candidates. While SWNT are very curly on production in general, MWNT are quite straight
and come in aligned arrays,
which is a great technological advantage in view of channeling application.

With high energy beams and typically short nanotubes available, the issue of
{\it dechanneling length} due to scattering on electrons within the bulk of the tube is less important
at the time. However, this issue is physically very interesting because of low electronic density
(and hence low electronic scattering) in the bulk of the nanotube, and it deserves a special study.

Whereas the interest to nanotube channeling at this early stage is more academic, some application can be quoted
in relation with unique capabilities of nanotube channeling. There is a need in beams of very small emittance,
"microbeams" or even "nanobeams", and channeling technique can be a help here \cite{23}.
The capability to produce beams with very small cross-section, i.e. "nano-beams", will be helpful
in medical, biological, and technological applications.
We refer to \cite{23,42} for discussion of it.

\section{Summary}

Channeling and centrifugal dechanneling phenomena have been demonstrated and
studied for positively charged particles in MWNT.
As shown in computer simulations, MWNT can channel particle beams with efficiency similar to that of
crystal channeling, and to that of SWNT, despite the fact that
the geometry of MWNT, and respectively the potential wells where channeled particles can be trapped, are
very different from the case of SWNT.

Together with the advantage of MWNT produced typically quite straight (unlike SWNT with strong parasitic
curvature) and in aligned arrays, this finding suggests that a MWNT is a better candidate for channeling than a SWNT.
Multi-wall nanotubes could make a basis for an efficient technique of beam steering at particle
accelerators.

\section*{Acknowledgements}

This work was partially supported by the Italian
Research Ministry MIUR, National Interest Program under grant
COFIN 2002022534, by INFN-International Affairs Funds, and by
INTAS-CERN Grants No. 132-2000 and 03-52-6155.

\pagebreak

\parindent=0pt

\section*{Figure captions}
\vspace{10mm}

{\bf Figure 1}
\vspace{3mm}

An example of the trajectory of a 1-GeV proton  injected between two walls
of a straight MWNT (a);
the same in a slightly bent (b), R = 14 cm, and strongly bent (c), R = 3 cm, MWNT.
\vspace{10mm}

{\bf Figure 2}
\vspace{3mm}

An example of 1-GeV proton angular distribution downstream of a bent MWNT, in the direction of bending.

\vspace{10mm}

{\bf Figure 3}
\vspace{3mm}

The number of 1-GeV protons channeled through a bent MWNT (50 $\mu$m long) shown as a function of the
centrifugal force $pv/R$; for comparison, also shown is the same dependence for SWNT (diameters 0.55 nm and 11 nm)
and for Si (110) crystal lattice.
Two ranges were studied: moderate bendings (a) and strong bendings (b).

\end{document}